\documentclass[%
 reprint,
superscriptaddress,
 amsmath,amssymb,
 aps,
 showkeys,
]{revtex4-2}

\usepackage{graphicx}% Include figure files
\usepackage{dcolumn}% Align table columns on decimal point
\usepackage{bm}% bold math
\usepackage{hyperref}
\usepackage[bottom]{footmisc}
\usepackage{footnote}
\usepackage{xcolor}
\hypersetup{breaklinks = true, colorlinks = true, citecolor = blue, linkcolor = blue, urlcolor = blue}

\usepackage[mathlines]{lineno}% Enable numbering of text and display math
\newcommand{\be}{\begin{equation}}
\newcommand{\ee}{\end{equation}}
\newcommand{\bea}{\begin{eqnarray}}
\newcommand{\eea}{\end{eqnarray}}

\newcommand{\AuAu}{Au$+$Au~}
\newcommand{\jpsi}{$J/\psi~$}
\newcommand{\ptSquare}{$p^{2}_{\mathrm{T}}~$}

\newcommand{\wGammaN}{$\left \langle W_{\mathrm{\gamma*N}}\right \rangle~$}
\newcommand{\sNNrhic}{$\sqrt{s_{_\mathrm{NN}}}=200$}

\begin{document}

\title{Observation of strong nuclear suppression in exclusive $J/\psi$ photoproduction \\in Au$+$Au ultra-peripheral collisions at RHIC}% Force line breaks with \\
\affiliation{Abilene Christian University, Abilene, Texas   79699}
\affiliation{AGH University of Krakow, FPACS, Cracow 30-059, Poland}
\affiliation{Argonne National Laboratory, Argonne, Illinois 60439}
\affiliation{American University in Cairo, New Cairo 11835, Egypt}
\affiliation{Ball State University, Muncie, Indiana, 47306}
\affiliation{Brookhaven National Laboratory, Upton, New York 11973}
\affiliation{University of Calabria \& INFN-Cosenza, Rende 87036, Italy}
\affiliation{University of California, Berkeley, California 94720}
\affiliation{University of California, Davis, California 95616}
\affiliation{University of California, Los Angeles, California 90095}
\affiliation{University of California, Riverside, California 92521}
\affiliation{Central China Normal University, Wuhan, Hubei 430079 }
\affiliation{University of Illinois at Chicago, Chicago, Illinois 60607}
\affiliation{Chongqing University, Chongqing, 401331}
\affiliation{Creighton University, Omaha, Nebraska 68178}
\affiliation{Czech Technical University in Prague, FNSPE, Prague 115 19, Czech Republic}
\affiliation{Technische Universit\"at Darmstadt, Darmstadt 64289, Germany}
\affiliation{National Institute of Technology Durgapur, Durgapur - 713209, India}
\affiliation{ELTE E\"otv\"os Lor\'and University, Budapest, Hungary H-1117}
\affiliation{Frankfurt Institute for Advanced Studies FIAS, Frankfurt 60438, Germany}
\affiliation{Fudan University, Shanghai, 200433 }
\affiliation{Guangxi Normal University, Guilin, 541004}
\affiliation{University of Heidelberg, Heidelberg 69120, Germany }
\affiliation{University of Houston, Houston, Texas 77204}
\affiliation{Huzhou University, Huzhou, Zhejiang  313000}
\affiliation{Indian Institute of Science Education and Research (IISER), Berhampur 760010 , India}
\affiliation{Indian Institute of Science Education and Research (IISER) Tirupati, Tirupati 517507, India}
\affiliation{Indian Institute Technology, Patna, Bihar 801106, India}
\affiliation{Indiana University, Bloomington, Indiana 47408}
\affiliation{Institute of Modern Physics, Chinese Academy of Sciences, Lanzhou, Gansu 730000 }
\affiliation{University of Jammu, Jammu 180001, India}
\affiliation{Kent State University, Kent, Ohio 44242}
\affiliation{University of Kentucky, Lexington, Kentucky 40506-0055}
\affiliation{Lawrence Berkeley National Laboratory, Berkeley, California 94720}
\affiliation{Lehigh University, Bethlehem, Pennsylvania 18015}
\affiliation{Max-Planck-Institut f\"ur Physik, Munich 80805, Germany}
\affiliation{Michigan State University, East Lansing, Michigan 48824}
\affiliation{National Institute of Science Education and Research, HBNI, Jatni 752050, India}
\affiliation{National Cheng Kung University, Tainan 70101 }
\affiliation{Nuclear Physics Institute of the CAS, Rez 250 68, Czech Republic}
\affiliation{The Ohio State University, Columbus, Ohio 43210}
\affiliation{Institute of Nuclear Physics PAN, Cracow 31-342, Poland}
\affiliation{Panjab University, Chandigarh 160014, India}
\affiliation{Purdue University, West Lafayette, Indiana 47907}
\affiliation{Rice University, Houston, Texas 77251}
\affiliation{Rutgers University, Piscataway, New Jersey 08854}
\affiliation{University of Science and Technology of China, Hefei, Anhui 230026}
\affiliation{South China Normal University, Guangzhou, Guangdong 510631}
\affiliation{Sejong University, Seoul, 05006, South Korea}
\affiliation{Shandong University, Qingdao, Shandong 266237}
\affiliation{Shanghai Institute of Applied Physics, Chinese Academy of Sciences, Shanghai 201800}
\affiliation{Southern Connecticut State University, New Haven, Connecticut 06515}
\affiliation{State University of New York, Stony Brook, New York 11794}
\affiliation{Instituto de Alta Investigaci\'on, Universidad de Tarapac\'a, Arica 1000000, Chile}
\affiliation{Temple University, Philadelphia, Pennsylvania 19122}
\affiliation{Texas A\&M University, College Station, Texas 77843}
\affiliation{University of Texas, Austin, Texas 78712}
\affiliation{Tsinghua University, Beijing 100084}
\affiliation{University of Tsukuba, Tsukuba, Ibaraki 305-8571, Japan}
\affiliation{University of Chinese Academy of Sciences, Beijing, 101408}
\affiliation{United States Naval Academy, Annapolis, Maryland 21402}
\affiliation{Valparaiso University, Valparaiso, Indiana 46383}
\affiliation{Variable Energy Cyclotron Centre, Kolkata 700064, India}
\affiliation{Warsaw University of Technology, Warsaw 00-661, Poland}
\affiliation{Wayne State University, Detroit, Michigan 48201}
\affiliation{Wuhan University of Science and Technology, Wuhan, Hubei 430065}
\affiliation{Yale University, New Haven, Connecticut 06520}

\author{M.~I.~Abdulhamid}\affiliation{American University in Cairo, New Cairo 11835, Egypt}
\author{B.~E.~Aboona}\affiliation{Texas A\&M University, College Station, Texas 77843}
\author{J.~Adam}\affiliation{Czech Technical University in Prague, FNSPE, Prague 115 19, Czech Republic}
\author{L.~Adamczyk}\affiliation{AGH University of Krakow, FPACS, Cracow 30-059, Poland}
\author{J.~R.~Adams}\affiliation{The Ohio State University, Columbus, Ohio 43210}
\author{I.~Aggarwal}\affiliation{Panjab University, Chandigarh 160014, India}
\author{M.~M.~Aggarwal}\affiliation{Panjab University, Chandigarh 160014, India}
\author{Z.~Ahammed}\affiliation{Variable Energy Cyclotron Centre, Kolkata 700064, India}
\author{E.~C.~Aschenauer}\affiliation{Brookhaven National Laboratory, Upton, New York 11973}
\author{S.~Aslam}\affiliation{Indian Institute Technology, Patna, Bihar 801106, India}
% ...

\collaboration{STAR Collaboration}\noaffiliation
 %STAR full author list, update before CWR/sumissions.

\author{STAR Collaboration}

\date{\today}% It is always \today, today,
\begin{abstract} 
We report a measurement of exclusive \jpsi and $\psi(2s)$ photoproduction in Au$+$Au ultra-peripheral collisions at $\sqrt{s_{_\mathrm{NN}}}=200$ GeV using the STAR detector. For the first time, i) the $\psi(2s)$ photoproduction in midrapidity at the Relativistic Heavy-Ion Collider has been experimentally measured; ii) nuclear suppression factors are measured for both the coherent and incoherent \jpsi production. At average photon-nucleon center-of-mass energy of 25.0 GeV, the coherent and incoherent \jpsi cross sections of Au nuclei are found to be $71\pm10\%$ and $36\pm7\%$, respectively, of that of free protons. The stronger suppression observed in the incoherent production provides a new experimental handle to study the initial-state parton density in heavy nuclei. Data are compared with theoretical models quantitatively. 
\end{abstract}

\keywords{ultra-peripheral collision, vector meson production, nuclear parton modification}%Use showkeys class option if keyword display desired
\maketitle
The fundamental structure of protons and neutrons, collectively known as nucleons, is at the core of understanding modern physics. They are directly connected to problems of color confinement, the microscopic structures of visible matter, and the origin of dynamical mass generation from nonperturbative Quantum Chromodynamics (QCD).
These problems are even more complex in the nuclear environment. Quark and gluon distributions
for bound nucleons inside nuclei could be
drastically different from those of the free nucleon. Understanding the fundamental structures of both
nucleons and nuclei in a consistent framework is one of the most pressing tasks in high energy nuclear physics.

In recent years, vector meson photoproduction in ultra-peripheral collisions (UPC) of heavy ions has 
provided an excellent experimental probe to study the structures of nucleons and nuclei~\cite{Arslandok:2023utm}. Typically, these photon-induced interactions take place at a large impact parameter and only produce one particle, e.g., the $J/\psi$ meson~\cite{Bertulani:2005ru}. In this reaction, the target nucleus may stay intact (coherent) or break up (incoherent), largely depending on the momentum transfer of the interaction. 

Specifically, coherent \jpsi photoproduction has been extensively investigated by heavy-ion collider experiments~\cite{PHENIX:2009xtn, Khachatryan:2016qhq,Abelev:2012ba,ALICE:2020ugp,ALICE:2021jnv,ALICE:2021tyx,ALICE:2021gpt,LHCb:2021hoq,CMS:2023snh,ALICE:2023jgu}, where the resulting cross sections are found to be significantly suppressed with respect to those of a free proton~\cite{ALICE:2012yye,Khachatryan:2016qhq,Abelev:2012ba,ALICE:2023jgu,CMS:2023snh}. Many models attempt to explain this phenomenon~\cite{Strikman:2018mbu,Guzey:2013qza,Guzey:2018tlk,Mantysaari:2022sux,Sambasivam:2019gdd, Toll:2012mb}, but the underlying mechanism remains highly debated~\cite{Arslandok:2023utm}. On the other hand, the nuclear suppression of incoherent vector meson production has never been explicitly measured. Comparing incoherent vector meson production on a heavy nucleus and on a free proton 
is equivalent to comparing the parton structure of bound and free nucleons. This is one of the most direct and unambiguous approaches for studying bound nucleons in heavy-ion collisions. 

In parallel to UPC measurements, hadronic proton-nucleus ($p+$A) collision data have been an important experimental handle on the nuclear parton densities in heavy nuclei~\cite{Armesto:2018ljh}. Despite a similar fundamental problem regarding the nuclear modification on parton distribution functions (nPDFs), UPC \jpsi and other vector mesons have the unique advantages of having little to no Multiple Parton Interaction~\cite{Sjostrand:2017cdm}, no radiative energy loss~\cite{Arleo:2021bpv}, and a well-controlled event topology and environment. These effects, on the other hand, may complicate the interpretation of $p+$A data in terms of nPDFs~\cite{Armesto:2018ljh}, e.g., charged hadron production, heavy-flavor production, jet productions, etc. Therefore, UPC measurements in this Letter provide important and complementary data to the modification of nuclear parton densities in heavy nuclei. Although the hard scale of the UPC process is mainly determined by the vector meson mass (or more precisely the quark mass), the data of \jpsi and $\psi(2s)$ in this report provide a unique constraint on the nPDFs at the fixed scale. However, at the forthcoming Electron-Ion Collider, the precise control of the photon virtuality, thus an independent hard scale, will further improve the constraints of nPDFs from vector meson production.

% Comment out after referee report - Mar 28, 2024
% In addition, the incoherent production is a direct measure of nucleon and parton density fluctuation based on the Good-Walker paradigm~\cite{PhysRev.120.1857}, which provides a dynamical picture of the structure of nuclei at high energy. 

In this Letter, we report measurements in \AuAu UPCs at \sNNrhic~GeV using the STAR detector at the Relativistic Heavy-Ion Collider (RHIC). Specifically, we measure: i) coherent and incoherent \jpsi photoproduction cross sections on Au nuclei, associated with different neutron emission patterns as detected in zero degree calorimeters (ZDCs); ii) photoproduction of $\psi(2s)$ at midrapidity; iii) nuclear suppression factors with respect to free nucleons. The average photon-nucleon center-of-mass energy, $\left \langle W_{\mathrm{\gamma*N}}\right \rangle$, is approximately 25.0 GeV for both coherent and incoherent processes at midrapidity in Au$+$Au UPCs~\footnote{Note that there is a $-$0.5 GeV shift in the estimate of \wGammaN at midrapidity $|y|<0.2$, caused by the higher photon flux of the lower energy photon contribution; however, the effect of this shift is found to be negligible.}. The nuclear suppression data are compared with theoretical models: the nuclear shadowing model with leading twist approximation (LTA)~\cite{Kryshen:2023bxy,Guzey:2013xba} and the saturation model color glass condensate (CGC)~\cite{Mantysaari:2022sux,Mantysaari:2016ykx}.

% detector
The Solenoidal Tracker at RHIC (STAR) detector~\cite{Ackermann:2002ad} and its subsystems have been thoroughly described in previous STAR papers~\cite{Adam:2018tdm,Adam:2020cwy}. Charged particle tracking, including transverse momentum reconstruction and charge sign determination, is provided by the Time Projection Chamber (TPC)~\cite{Anderson:2003ur} positioned in a 0.5 T solenoidal magnetic field. The TPC volume extends from 50 to 200 cm from the beam axis and covers pseudorapidities $|\eta|<1.0$ over the full azimuthal angle, $0<\phi<2\pi$.
The TPC also provides ionization loss ($dE/dx$) measurement of tracks used for particle identification.
Surrounding the TPC is the Barrel Electromagnetic Calorimeter (BEMC)~\cite{Beddo:2002zx}, which is a
lead-scintillator sampling calorimeter. The BEMC is segmented into 4800 optically isolated
towers covering the full azimuthal angle for pseudorapidities $|\eta|<1.0$.
Between the TPC and BEMC is the Time Of Flight (TOF) system. It is finely segmented in $\eta$ and $\phi$ and provides fast trigger signals for charged particles in the range $|\eta| < 0.9$.
There are two Beam-Beam Counters (BBCs)~\cite{Whitten:2008zz}, one on each side of the STAR main detector, covering a pseudorapidity range of $3.4<|\eta|<5.0$. There are also two ZDCs~\cite{Ackermann:2002ad},
18 m along each beam direction covering $|\eta| > 6.7$,
used to monitor the
luminosity and to tag forward neutrons.

% data
The UPC data were collected by the STAR experiment during the 2016 \AuAu run, which corresponds to an integrated luminosity of 13.5~$\rm nb^{-1}$ yielding approximately $2.4\times10^7$ UPC \jpsi triggered events. The \jpsi and $\psi(2s)$ candidates are reconstructed via the electron decay channel, \jpsi$ (\psi(2s)) \rightarrow e^{+}e^{-}$. Based on this channel, the UPC \jpsi ($\psi(2s)$) trigger is defined by 
a topological selection of back-to-back clusters in the BEMC,
a TOF charged track multiplicity between 2 and 6 (inclusive),
and no BBC signal in either beam direction.

In the offline analysis, the events are required to have a pair of tracks with a vertex that is reconstructed within 100 cm of the center of the STAR detector along the beam direction.
Tracks are required to have at least 15 points (out of a maximum of 45) to ensure sufficient momentum resolution, contain no fewer than 11 points for the ionization energy loss determination to ensure good $dE/dx$ resolution, and to be matched to a BEMC cluster for consistency with the trigger.
Electron pair selection is performed based on the $dE/dx$ of tracks, where the dominant contamination at $p_{\rm T}~\sim~1.5~\rm{GeV/c}$ is from pions. The variable $n_{\sigma,e}$ ($n_{\sigma,\pi}$) is the difference between the measured $dE/dx$ value compared to an electron ($\pi$) hypothesis of the predicted $dE/dx$ value.  It is calculated in terms of the number of standard deviations from the predicted mean. The pair selection variable $\chi^{2}_{ee}$ is defined as $n^{2}_{\sigma,e1}+n^{2}_{\sigma,e2}$  for tracks 1 and 2, and similarly for the $\pi$ pair hypothesis.
Tracks consistent with electron pairs were selected by requiring $\chi^2_{ee}<10$ and those consistent with pion pairs were rejected by requiring $\chi^2_{ee} < \chi^2_{\pi\pi}$.
The selections were performed separately for opposite-sign ($+-$) and like-sign (++,$--$) pairs.
The like-sign pairs were taken as a measure of
combinatoric backgrounds and subtracted from the
opposite-sign pairs for final distributions.

Like-sign subtracted distributions of
invariant mass $m_{ee}$ and pair $p_{\rm T}$
were produced.
Template distributions of signal \jpsi and 
background $e^+e^-$ from Quantum Electrodynamics (QED) photon-photon interactions
and $\psi(2S)$ production were also created.
The templates used output from the
STARlight~\cite{Klein:2016yzr} Monte Carlo program weighted by the H1 data~\cite{Alexa:2013xxa}, which were passed through the GEANT3-based~\cite{Brun:1987ma} STAR detector simulation to model the detector response.
To extract the \jpsi and $\psi(2s)$ yield, simultaneous fits of 
the templates to both the $m_{ee}$ and pair $p_{\rm T}$ distributions were performed.
% The raw \jpsi yield of the entire analyzed sample after full event selections and background subtraction is $\sim7900$. 
For the differential cross section measurements in rapidity intervals, the same procedure was applied, and the \jpsi yields
were extracted for both coherent and incoherent production. See the detailed procedure in Ref.~\cite{STAR:2023gpk}.

% neutron classes
The results were further divided into different
neutron emission patterns as measured by the ZDCs, where neutrons can be produced by either the QED process of mutual Coulomb excitation~\cite{Bertulani:2005ru} or nuclear breakup from hard scattering processes.
%Similar to Ref.~\cite{STAR:2021wwq}, DONE FOR dAu?
The patterns of neutron emission are categorized as:
i) $0n0n$ - neither ZDC has detected a neutron;
ii) $0nXn$ - one ZDC has detected at least one neutron and the other has no neutrons;
iii) $XnXn$ - both ZDCs have detected at least one neutron. Results summing over these three categories are
denoted as the $all~n$ category.
Overlaps of ZDC hits from other events in the same
RHIC bunch crossing caused migrations between these categories.
This effect was measured in a sample of zero-bias data in terms of ZDC requirement,
and the migrations were corrected for.
This correction was up to 8\% for the coherent \jpsi photoproduction cross section.

\begin{figure}[thb]
\includegraphics[width=3.4in]{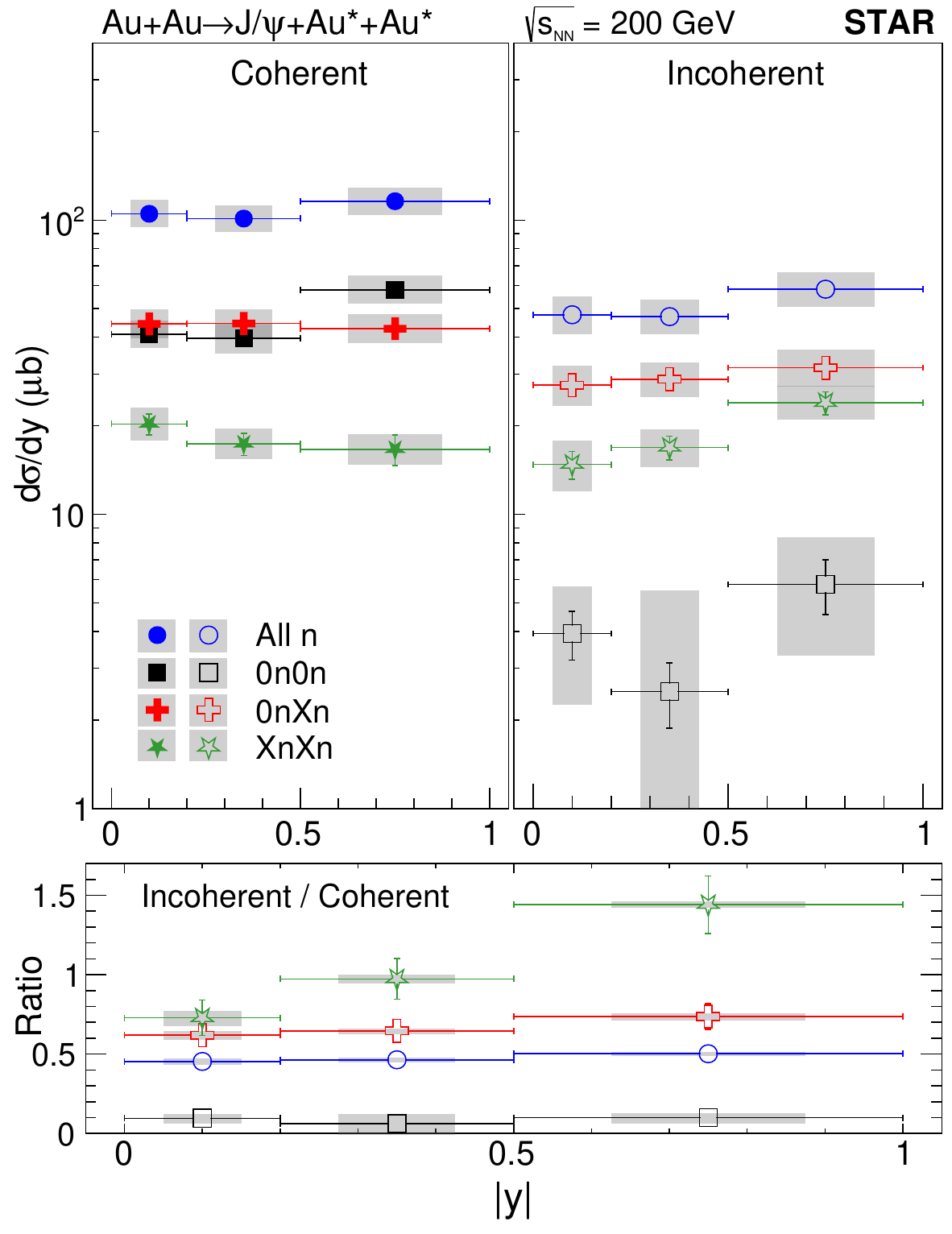}
  \caption{ \label{fig:res:figure_2} Differential cross sections $d\sigma/dy$ for coherent (top left) and incoherent (top right) \jpsi photoproduction and their ratios (bottom) as a function of $|y|$ in \AuAu UPCs at \sNNrhic~GeV,
  for the different neutron categories defined in the text.
  Statistical uncertainties are represented by the error bars, and the systematic uncertainties are denoted as boxes.
  There is a systematic uncertainty on the cross sections of 10\% from the integrated luminosity that is not shown.}
\end{figure}

The differential cross section of \jpsi photoproduction as a function of \ptSquare and rapidity $y$ is measured in the \AuAu UPCs as follows, 
\begin{align}
	\frac{d^{2}\sigma}{dp^{2}_{T}dy}_i =
    \frac{N_{raw,i}}{\varepsilon_{trig} \cdot corr_i \cdot L \cdot BR \cdot \Delta p^{2}_{Ti} \cdot 2\Delta y_i } \, .
    \label{eq:crosssec}
\end{align}
Here $\frac{d^{2}\sigma}{dp^{2}_{T}dy}_i$ is the doubly differential cross section in $(p^{2}_T,y)$ bin $i$, where $i$ is a single index that includes all measured $(p^{2}_T,y)$ combinations.
\noindent $N_{raw,i}$ is the raw number of \jpsi in bin $i$. $\varepsilon_{trig}$ is an overall scale correction for trigger efficiency, and $corr_i$ is the acceptance and efficiency correction for bin $i$. The integrated luminosity is denoted by $L$, and $BR=5.97\%$ is the branching ratio of \jpsi decaying into an electron and positron~\cite{ParticleDataGroup:2020ssz}; $\Delta p^{2}_{Ti}$ and $\Delta y_i$ are the widths
of bin $i$.
The factor of 2 is introduced because \jpsi events are measured within $|y|<1$, while the cross section
is only reported for $y>0$;
this factor is not included for the cross section
$d\sigma/dy$ for $0nXn$ discussed later.
The \jpsi acceptance corrections are based on the STARlight~\cite{Klein:2016yzr} MC events embedded into STAR zero-bias events, where a bin-by-bin unfolding technique is employed in the correction procedure as exhibited in Eq.~\ref{eq:crosssec}.

For each rapidity of $J/\psi$ there are two different contributions mixed together, a higher energy and a lower energy photon, which correspond to photons emitted by either nucleus. Based on the technique in Ref.~\cite{Guzey:2013jaa}, we resolve the photon energy ambiguity and measure the coherent \jpsi photoproduction cross section for $\rm \gamma+Au\rightarrow J/\psi+Au$. The details of this procedure are outlined in the article Ref.~\cite{STAR:2023gpk} submitted along with this Letter. 

Different sources of systematic uncertainty on the differential cross section were investigated, which are similar to the previous STAR publication on \jpsi photoproduction in deuterons~\cite{STAR:2021wwq}. The $\psi(2s)$ production shares similar systematic uncertainties, except for the feed-down correction from $\psi(2s)\rightarrow J/\psi + X$. The background subtraction using fit templates introduces uncertainties from the fitting, resulting in 10-20\% on the background-subtracted distributions, depending on the \jpsi transverse momentum.
Several factors contribute to the acceptance and efficiency corrections for pair mass and $p_T$ distributions. The trigger efficiency determination results in final uncertainties $\sim8\%$.
The efficiency of matching tracks to BEMC energy deposits as measured with data has an uncertainty of $\sim5\%$ on the pair efficiency. 
The uncertainty on weighting of STARlight to match the $p_T$ distributions is only significant on the steeply falling coherent \jpsi peak, where the pair detection efficiency uncertainty is up to 15\%. 
The uncertainty from modeling radiative events in the simulation is $\sim 2\%$ on pair acceptance. The background subtraction and acceptance uncertainties, including feed-down corrections and branching ratio, were determined bin-by-bin in mass and $p_{\rm T}$ of electron pairs. 
They were added in quadrature along with an overall 4\% uncertainty on track and vertex reconstruction efficiency; this sum is shown with the displayed data points. The systematic uncertainty on modeling the transversely polarized photon flux is found to be up to 3.5\% by varying the Au radius $\pm 0.5~\rm{fm}$, where the same method has been adopted as in Ref.~\cite{STAR:2021wwq}. Finally, there is an uncertainty of 10\% on the luminosity measurement, which is the dominant systematic uncertainty source, resulting in a scale uncertainty of 10\% on all cross sections, which is not displayed in the figures.

In Fig.~\ref{fig:res:figure_2}, the differential cross sections $d\sigma/dy$ of \jpsi photoproduction as a function of $|y|$ for coherent (left) and incoherent (right) production are presented, for $all~n$ data and each neutron category separately.
They are obtained by integrating the data over low (coherent) or high (incoherent) $p^{2}_{\rm{T}}$ and using the template fits to correct to the full $p^{2}_{\rm{T}}$ range~\cite{STAR:2023gpk}. The coherent spectra dominate for $p^{2}_{\rm{T}}<0.02~\rm{(GeV/c)^2}$, while incoherent ones dominate the higher  $p^{2}_{\rm{T}}$. 
The rapidity interval includes both positive and negative rapidities, where the data are plotted at the bin center. The ratio between incoherent and coherent \jpsi production is also shown in the bottom panel. Note that the $0n0n$ in incoherent production is mostly dominated by non-neutron breakup from the target nucleus~\cite{Chang:2021jnu}.

\begin{figure}[thb]
\includegraphics[width=3.4in]{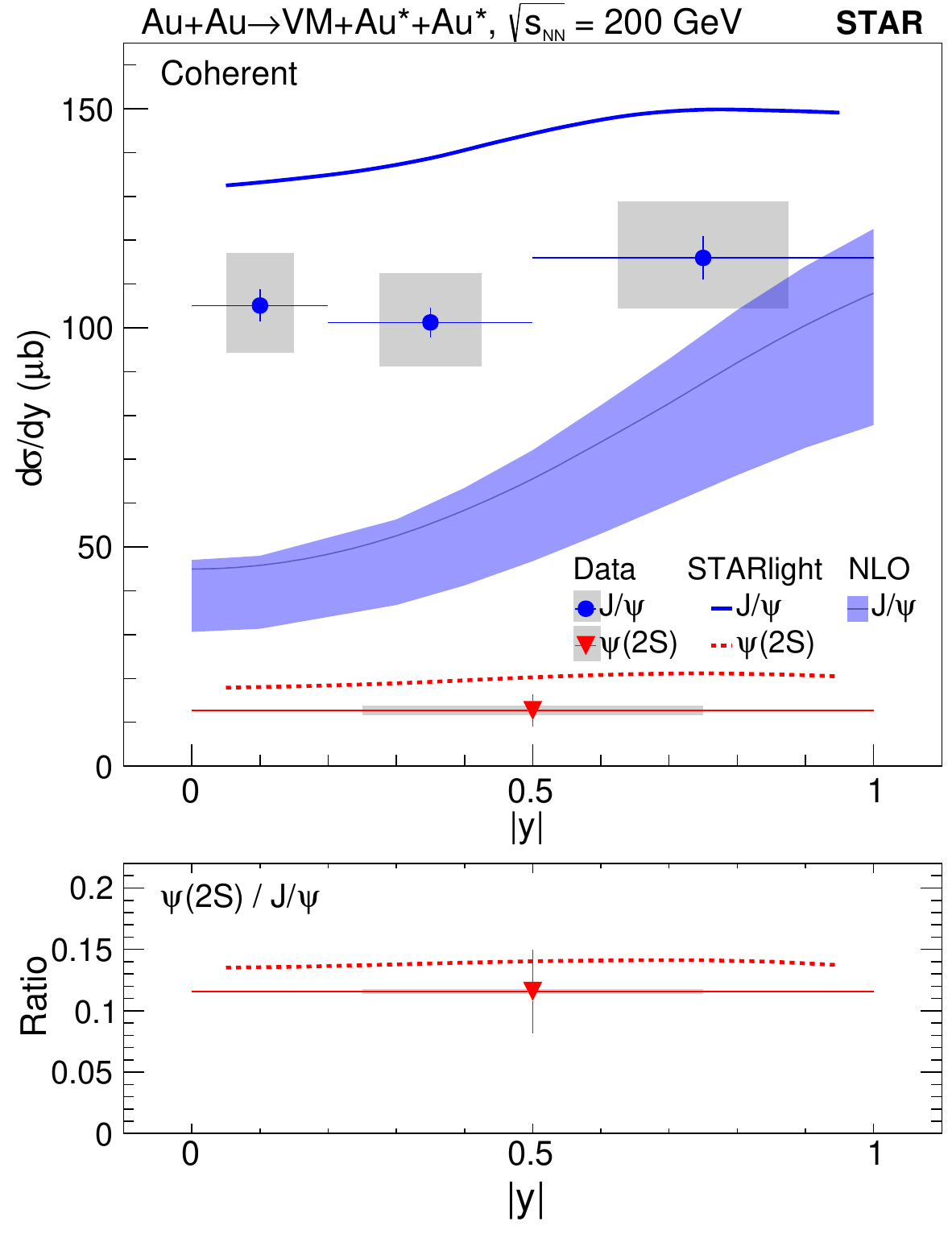}
  \caption{ \label{fig:res:figure_8_b} Differential cross section $d\sigma/dy$ for coherent \jpsi and $\psi(2s)$ photoproduction as a function of $|y|$ in \AuAu UPCs at \sNNrhic~GeV.
  The STARlight model~\cite{Klein:2016yzr}
  and the NLO pQCD calculations~\cite{Eskola:2022vaf,Eskola:2022vpi}
  are compared with the data. Ratio between $\psi(2s)$ and \jpsi is shown in the bottom panel. Statistical uncertainty is represented by the error bars, and the systematic uncertainty is denoted as boxes. There is a systematic uncertainty of 10\% from the integrated luminosity that is not shown.  }
\end{figure}

In Fig.~\ref{fig:res:figure_8_b}, the first measurement of exclusive $\psi(2s)$ photoproduction has been reported at the top RHIC energy. The differential cross section $d\sigma/dy$ compares with that of the coherent \jpsi photoproduction and the STARLight model. The ratio between \jpsi and $\psi(2s)$ is shown in the bottom panel, where most systematic uncertainties are canceled. The ratio is found to be similar to the same measurement at 5.02 TeV at the LHC~\cite{ALICE:2021gpt}, and consistent with most models and data for $pp$ or $ep$ collisions~\cite{Ducati:2013bya,H1:2002yab,CDF:2009xey,LHCb:2014acg}. Within the uncertainty of our measurement, we did not see any significantly different modification of $\psi(2s)$ relative to \jpsi in Au$+$Au UPCs at RHIC energies with respect to free protons. Moreover, the colored band in 
Fig.~\ref{fig:res:figure_8_b}
shows the first NLO perturbative QCD calculation of the \jpsi photoproduction at RHIC energies. The input nPDF is from EPPS21~\cite{Eskola:2021mjl}, where the current uncertainty coming from the nPDF on the \jpsi production cross section can be as large as 50\% to 160\%.
Consequently, this is not shown.
The uncertainty band shown on the figure is only based on the scale uncertainty. For details, see Refs~\cite{Eskola:2022vaf,Eskola:2022vpi}. This prediction has been found to be underestimated by more than a factor of 2 at mid-rapidity and 10-20\% at higher rapidity. This data will significantly constrain the nPDF at the NLO for both quarks and gluons.

% To further investigate the rapidity dependence, the rapidity distributions for coherent and incoherent \jpsi
% photoproduction are shown for the asymmetric $0nXn$ neutron category in Fig.~\ref{fig:res:figure_3}. Positive \jpsi rapidity ($\rm{sgn(y_{n})\cdot y_{J/\psi}}$) is defined by the direction of forward going neutrons, $\rm{sgn(y_{n})}$, where $\rm{y_{ n}}$ is the neutron rapidity.
% The coherent \jpsi rapidity distribution
% is found to be symmetric under the transformation $y\rightarrow -y$. Neutron emission for coherent \jpsi photoproduction occurs through mutual Coulomb excitation, in which the neutron may be emitted by either nucleus. Thus, the neutron direction is not expected to be correlated to the \jpsi direction. Before this measurement, the assumption that neutron emission is independent of coherent vector meson photoproduction has never been experimentally tested.
% By contrast, in the incoherent process the target nucleus breaks up in the hard interaction, and neutrons hitting a ZDC identify the direction
% of the target nucleus; for the $0nXn$ configuration, this direction is unambiguous. This result is also compared with an incoherent \jpsi production model, BeAGLE~\cite{Chang:2022hkt}, which is found to be qualitatively consistent with the data. 

\begin{figure}[thb]
\includegraphics[width=3.4in]{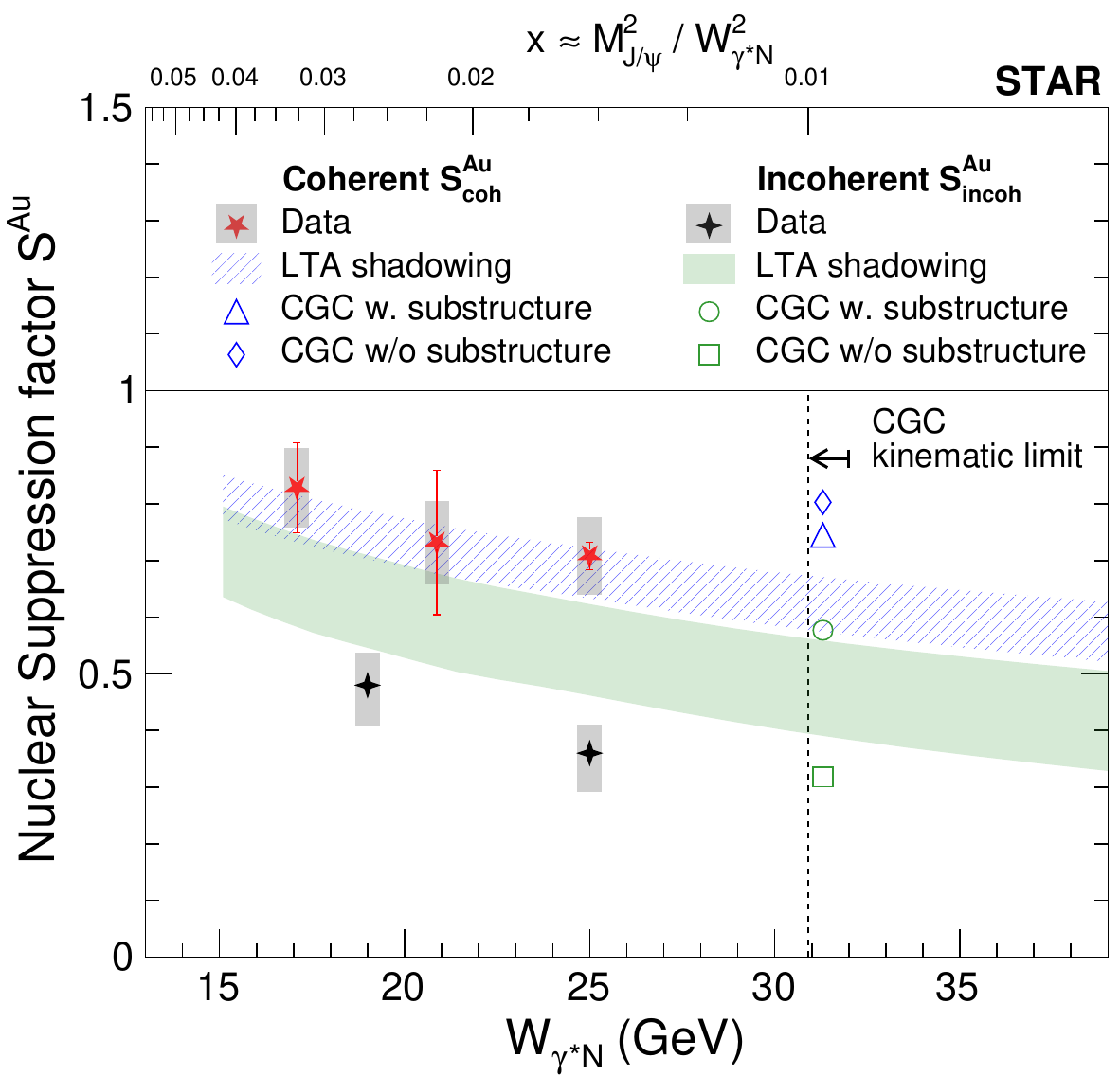}
  \caption{ \label{fig:res:figure_4} Nuclear suppression factor of coherent ($\rm{S^{Au}_{coh}}$, with respect to the Impulse Approximation) and incoherent ($\rm{S^{Au}_{incoh}}$, with respect to the HERA data~\cite{Alexa:2013xxa}) \jpsi photoproduction in Au$+$Au UPCs. The data are compared with the nuclear shadowing model~\cite{Kryshen:2023bxy} and the CGC model~\cite{Mantysaari:2022sux}. The CGC points are shifted from the vertical line for better visibility. Statistical uncertainties are represented by the error bars, and the systematic uncertainties are denoted as boxes. There is a systematic uncertainty of 10\% from the integrated luminosity that is not shown. 
 }
\end{figure}

In Fig.~\ref{fig:res:figure_4}, the nuclear suppression factors of coherent ($\rm{S^{Au}_{coh}}$) and incoherent ($\rm{S^{Au}_{incoh}}$) \jpsi photoproduction are shown as a function of \wGammaN in Au$+$Au UPCs. For the coherent case, the $\rm S^{Au}_{coh}$ is calculated based on the ratio between the coherent \jpsi cross section of $all~n$ and the Impulse Approximation (IA)~\cite{Guzey:2013xba}, where IA represents the scenario without any nuclear effect constrained by free proton data~\cite{Guzey:2013xba}. The suppression factor at \wGammaN$=25.0~\rm{GeV}$ is found to be $0.71\pm0.04~\pm0.07\pm0.07$. The first quoted error is the model uncertainty on IA~\cite{Guzey:2013xba} for Au nuclei and the second error is a combination of statistics and systematic uncertainties added in quadrature, while the third is from the scale
uncertainty of the integrated luminosity. Note that the data, LTA, and CGC calculations use the same IA calculation to ensure proper comparisons. Furthermore, since the estimate of IA of $\psi(2s)$ is less constrained than that of $J/\psi$~\cite{H1:2002yab}, the corresponding nuclear suppression factor is not reported here. 

For the incoherent suppression factor, $\rm S^{Au}_{incoh}$, it is defined as the ratio between the incoherent \jpsi cross section of $all~n$ and the free proton data at HERA. In order to compare with photoproduction in $ep$ collisions, we use the published H1 data and its well-constrained parametrization~\cite{Alexa:2013xxa}. It is found that the STAR UPC incoherent \ptSquare distribution is well described by the H1 $ep$ template, with a suppression factor found to be $0.36^{+0.03}_{-0.04}\pm0.04\pm0.04$ at \wGammaN$=25.0~\rm{GeV}$. Here the first uncertainty is the H1 parametrization uncertainty, the second one is from the measurement that includes statistical and systematic uncertainty, and the third is the scale uncertainty on the integrated luminosity. 
The details of this procedure, both for coherent and incoherent processes, are outlined in the article Ref.~\cite{STAR:2023gpk} submitted along with this Letter. 

The nuclear shadowing model LTA and the saturation model CGC are compared with the data quantitatively. For the LTA, the upper bound of each band is showing the weak shadowing mode, while the lower bound shows the strong shadowing mode~\cite{Kryshen:2023bxy}. It is found that, for the first time, the incoherent suppression factor is less than that of the coherent production, as well as the strong shadowing mode in the LTA model. For the CGC model, although it is not strictly calculated at the STAR kinematic range due to the applicability of the model ($x>0.01$, where $x$ is the momentum fraction the parton carries of the nucleon), the incoherent data are found to be between the model scenarios calculated with or without sub-nucleonic fluctuation of the parton density~\cite{Mantysaari:2022sux}. Based on this data, it is hard to conclude if sub-nucleonic parton density fluctuation is present in the incoherent \jpsi photoproduction, contrary to the conclusion to a recent measurement by the ALICE Collaboration~\cite{ALICE:2023gcs}. Note that the \ptSquare distribution of the incoherent production are found to be consistent between STAR and ALICE. Nevertheless, the reported data provide new insights to the nuclear suppression in both coherent and incoherent \jpsi photoproduction, where incoherent production is found to be more suppressed than the coherent counterparts. Within the model of LTA and other models that are based on nuclear PDFs, this new data will significantly constrain the quantitative description of nuclear parton densities in heavy nuclei, complementary to other measurements in $p+$A collisions. 

% conclusion
In conclusion, differential cross sections of \jpsi photoproduction in coherent and incoherent processes as a function of rapidity $y$ and the first measurement of photoproduction of $\psi(2s)$ in \AuAu UPCs at \sNNrhic~GeV have been reported. These cross sections are measured separately in different neutron emission 
categories, as detected by the zero degree calorimeters. 
% It is observed that in the asymmetric neutron configuration the coherent \jpsi production is independent of the neutron emission direction, while the incoherent production has a strong dependence. This is consistent with expectations from Monte Carlo simulations, but confirmed experimentally for the first time. 
The relative coherent cross section to that of a free nucleon is found to be $71\pm10\%$ ($\sim30\%$ suppressed). The incoherent \jpsi photoproduction has been compared to that of a free proton based on the H1 data, where a stronger suppression than that of the coherent production is observed with a relative cross section of $36\pm7\%$ ($\sim60\%$ suppressed). This is stronger than predictions from the nuclear shadowing model, and does not directly support the CGC model with sub-nucleonic fluctuation.  The parton density at the top RHIC energy lies in the transition region ($x_{\rm{parton}}\sim~0.01$) between large momentum quarks ($x_{\rm{parton}}>0.1$) and low momentum gluons ($x_{\rm{parton}}<0.001$), which is essential to the understanding of both gluon saturation and nuclear shadowing mechanisms. This measurement provides important constraints to the parton density and is an essential experimental baseline for such measurements at the upcoming Electron-Ion Collider.

% Acknowledgement 
We thank the RHIC Operations Group and RCF at BNL, the NERSC Center at LBNL, and the Open Science Grid consortium for providing resources and support.  This work was supported in part by the Office of Nuclear Physics within the U.S. DOE Office of Science, the U.S. National Science Foundation, National Natural Science Foundation of China, Chinese Academy of Science, the Ministry of Science and Technology of China and the Chinese Ministry of Education, the Higher Education Sprout Project by Ministry of Education at NCKU, the National Research Foundation of Korea, Czech Science Foundation and Ministry of Education, Youth and Sports of the Czech Republic, Hungarian National Research, Development and Innovation Office, New National Excellency Programme of the Hungarian Ministry of Human Capacities, Department of Atomic Energy and Department of Science and Technology of the Government of India, the National Science Centre and WUT ID-UB of Poland, the Ministry of Science, Education and Sports of the Republic of Croatia, German Bundesministerium f\"ur Bildung, Wissenschaft, Forschung and Technologie (BMBF), Helmholtz Association, Ministry of Education, Culture, Sports, Science, and Technology (MEXT), Japan Society for the Promotion of Science (JSPS) and Agencia Nacional de Investigaci\'on y Desarrollo (ANID) of Chile.
\bibliography{reference}% Produces the bibliography via BibTeX.

\begin{thebibliography}{48}%
\makeatletter
\providecommand \@ifxundefined [1]{%
 \@ifx{#1\undefined}
}%
\providecommand \@ifnum [1]{%
 \ifnum #1\expandafter \@firstoftwo
 \else \expandafter \@secondoftwo
 \fi
}%
\providecommand \@ifx [1]{%
 \ifx #1\expandafter \@firstoftwo
 \else \expandafter \@secondoftwo
 \fi
}%
\providecommand \natexlab [1]{#1}%
\providecommand \enquote  [1]{``#1''}%
\providecommand \bibnamefont  [1]{#1}%
\providecommand \bibfnamefont [1]{#1}%
\providecommand \citenamefont [1]{#1}%
\providecommand \href@noop [0]{\@secondoftwo}%
\providecommand \href [0]{\begingroup \@sanitize@url \@href}%
\providecommand \@href[1]{\@@startlink{#1}\@@href}%
\providecommand \@@href[1]{\endgroup#1\@@endlink}%
\providecommand \@sanitize@url [0]{\catcode `\\12\catcode `\$12\catcode
  `\&12\catcode `\#12\catcode `\^12\catcode `\_12\catcode `\%12\relax}%
\providecommand \@@startlink[1]{}%
\providecommand \@@endlink[0]{}%
\providecommand \url  [0]{\begingroup\@sanitize@url \@url }%
\providecommand \@url [1]{\endgroup\@href {#1}{\urlprefix }}%
\providecommand \urlprefix  [0]{URL }%
\providecommand \Eprint [0]{\href }%
\providecommand \doibase [0]{https://doi.org/}%
\providecommand \selectlanguage [0]{\@gobble}%
\providecommand \bibinfo  [0]{\@secondoftwo}%
\providecommand \bibfield  [0]{\@secondoftwo}%
\providecommand \translation [1]{[#1]}%
\providecommand \BibitemOpen [0]{}%
\providecommand \bibitemStop [0]{}%
\providecommand \bibitemNoStop [0]{.\EOS\space}%
\providecommand \EOS [0]{\spacefactor3000\relax}%
\providecommand \BibitemShut  [1]{\csname bibitem#1\endcsname}%
\let\auto@bib@innerbib\@empty
%</preamble>
\bibitem [{\citenamefont {Arslandok}\ \emph {et~al.}(2023)\citenamefont
  {Arslandok} \emph {et~al.}}]{Arslandok:2023utm}%
  \BibitemOpen
  \bibfield  {author} {\bibinfo {author} {\bibfnamefont {M.}~\bibnamefont
  {Arslandok}} \emph {et~al.},\ }\bibfield  {title} {\bibinfo {title} {{Hot QCD
  White Paper}},\ }\href@noop {} {\  (\bibinfo {year} {2023})},\ \Eprint
  {https://arxiv.org/abs/2303.17254} {arXiv:2303.17254 [nucl-ex]} \BibitemShut
  {NoStop}%
\bibitem [{\citenamefont {Bertulani}\ \emph {et~al.}(2005)\citenamefont
  {Bertulani}, \citenamefont {Klein},\ and\ \citenamefont
  {Nystrand}}]{Bertulani:2005ru}%
  \BibitemOpen
  \bibfield  {author} {\bibinfo {author} {\bibfnamefont {C.~A.}\ \bibnamefont
  {Bertulani}}, \bibinfo {author} {\bibfnamefont {S.~R.}\ \bibnamefont
  {Klein}},\ and\ \bibinfo {author} {\bibfnamefont {J.}~\bibnamefont
  {Nystrand}},\ }\bibfield  {title} {\bibinfo {title} {{Physics of
  ultra-peripheral nuclear collisions}},\ }\href
  {https://doi.org/10.1146/annurev.nucl.55.090704.151526} {\bibfield  {journal}
  {\bibinfo  {journal} {Ann. Rev. Nucl. Part. Sci.}\ }\textbf {\bibinfo
  {volume} {55}},\ \bibinfo {pages} {271} (\bibinfo {year} {2005})},\ \Eprint
  {https://arxiv.org/abs/nucl-ex/0502005} {arXiv:nucl-ex/0502005} \BibitemShut
  {NoStop}%
\bibitem [{\citenamefont {Afanasiev}\ \emph {et~al.}(2009)\citenamefont
  {Afanasiev} \emph {et~al.}}]{PHENIX:2009xtn}%
  \BibitemOpen
  \bibfield  {author} {\bibinfo {author} {\bibfnamefont {S.}~\bibnamefont
  {Afanasiev}} \emph {et~al.} (\bibinfo {collaboration} {PHENIX}),\ }\bibfield
  {title} {\bibinfo {title} {{Photoproduction of J/psi and of high mass e+e- in
  ultra-peripheral Au+Au collisions at $\sqrt{s_{\mathrm{NN}}}$ = 200-GeV}},\
  }\href {https://doi.org/10.1016/j.physletb.2009.07.061} {\bibfield  {journal}
  {\bibinfo  {journal} {Phys. Lett. B}\ }\textbf {\bibinfo {volume} {679}},\
  \bibinfo {pages} {321} (\bibinfo {year} {2009})},\ \Eprint
  {https://arxiv.org/abs/0903.2041} {arXiv:0903.2041 [nucl-ex]} \BibitemShut
  {NoStop}%
\bibitem [{\citenamefont {Khachatryan}\ \emph {et~al.}(2017)\citenamefont
  {Khachatryan} \emph {et~al.}}]{Khachatryan:2016qhq}%
  \BibitemOpen
  \bibfield  {author} {\bibinfo {author} {\bibfnamefont {V.}~\bibnamefont
  {Khachatryan}} \emph {et~al.} (\bibinfo {collaboration} {CMS}),\ }\bibfield
  {title} {\bibinfo {title} {{Coherent $J/\psi$ photoproduction in
  ultra-peripheral PbPb collisions at $\sqrt{s_{\mathrm{NN}}} = $ 2.76 TeV with
  the CMS experiment}},\ }\href
  {https://doi.org/10.1016/j.physletb.2017.07.001} {\bibfield  {journal}
  {\bibinfo  {journal} {Phys. Lett. B}\ }\textbf {\bibinfo {volume} {772}},\
  \bibinfo {pages} {489} (\bibinfo {year} {2017})},\ \Eprint
  {https://arxiv.org/abs/1605.06966} {arXiv:1605.06966 [nucl-ex]} \BibitemShut
  {NoStop}%
\bibitem [{\citenamefont {Abelev}\ \emph
  {et~al.}(2013{\natexlab{a}})\citenamefont {Abelev} \emph
  {et~al.}}]{Abelev:2012ba}%
  \BibitemOpen
  \bibfield  {author} {\bibinfo {author} {\bibfnamefont {B.}~\bibnamefont
  {Abelev}} \emph {et~al.} (\bibinfo {collaboration} {ALICE}),\ }\bibfield
  {title} {\bibinfo {title} {{Coherent $J/\psi$ photoproduction in
  ultra-peripheral Pb-Pb collisions at $\sqrt{s_{\mathrm{NN}}} = 2.76$ TeV}},\
  }\href {https://doi.org/10.1016/j.physletb.2012.11.059} {\bibfield  {journal}
  {\bibinfo  {journal} {Phys. Lett. B}\ }\textbf {\bibinfo {volume} {718}},\
  \bibinfo {pages} {1273} (\bibinfo {year} {2013}{\natexlab{a}})},\ \Eprint
  {https://arxiv.org/abs/1209.3715} {arXiv:1209.3715 [nucl-ex]} \BibitemShut
  {NoStop}%
\bibitem [{\citenamefont {Acharya}\ \emph {et~al.}(2020)\citenamefont {Acharya}
  \emph {et~al.}}]{ALICE:2020ugp}%
  \BibitemOpen
  \bibfield  {author} {\bibinfo {author} {\bibfnamefont {S.}~\bibnamefont
  {Acharya}} \emph {et~al.} (\bibinfo {collaboration} {ALICE}),\ }\bibfield
  {title} {\bibinfo {title} {{Coherent photoproduction of $\rho^{0}$ vector
  mesons in ultra-peripheral Pb-Pb collisions at $\sqrt{s_{\mathrm{NN}}}$ =
  5.02 TeV}},\ }\href {https://doi.org/10.1007/JHEP06(2020)035} {\bibfield
  {journal} {\bibinfo  {journal} {JHEP}\ }\textbf {\bibinfo {volume} {06}},\
  \bibinfo {pages} {035 (2020)}},\ \Eprint {https://arxiv.org/abs/2002.10897}
  {arXiv:2002.10897 [nucl-ex]} \BibitemShut {NoStop}%
\bibitem [{\citenamefont {Acharya}\ \emph
  {et~al.}(2021{\natexlab{a}})\citenamefont {Acharya} \emph
  {et~al.}}]{ALICE:2021jnv}%
  \BibitemOpen
  \bibfield  {author} {\bibinfo {author} {\bibfnamefont {S.}~\bibnamefont
  {Acharya}} \emph {et~al.} (\bibinfo {collaboration} {ALICE}),\ }\bibfield
  {title} {\bibinfo {title} {{First measurement of coherent \ensuremath{\rho}0
  photoproduction in ultra-peripheral Xe\textendash{}Xe collisions at
  $\sqrt{s_{\mathrm{NN}}}$ = 5.44 TeV}},\ }\href
  {https://doi.org/10.1016/j.physletb.2021.136481} {\bibfield  {journal}
  {\bibinfo  {journal} {Phys. Lett. B}\ }\textbf {\bibinfo {volume} {820}},\
  \bibinfo {pages} {136481} (\bibinfo {year} {2021}{\natexlab{a}})},\ \Eprint
  {https://arxiv.org/abs/2101.02581} {arXiv:2101.02581 [nucl-ex]} \BibitemShut
  {NoStop}%
\bibitem [{\citenamefont {Acharya}\ \emph
  {et~al.}(2021{\natexlab{b}})\citenamefont {Acharya} \emph
  {et~al.}}]{ALICE:2021tyx}%
  \BibitemOpen
  \bibfield  {author} {\bibinfo {author} {\bibfnamefont {S.}~\bibnamefont
  {Acharya}} \emph {et~al.} (\bibinfo {collaboration} {ALICE}),\ }\bibfield
  {title} {\bibinfo {title} {{First measurement of the $|t|$-dependence of
  coherent $J/\psi$ photonuclear production}},\ }\href
  {https://doi.org/10.1016/j.physletb.2021.136280} {\bibfield  {journal}
  {\bibinfo  {journal} {Phys. Lett. B}\ }\textbf {\bibinfo {volume} {817}},\
  \bibinfo {pages} {136280} (\bibinfo {year} {2021}{\natexlab{b}})},\ \Eprint
  {https://arxiv.org/abs/2101.04623} {arXiv:2101.04623 [nucl-ex]} \BibitemShut
  {NoStop}%
\bibitem [{\citenamefont {Acharya}\ \emph
  {et~al.}(2021{\natexlab{c}})\citenamefont {Acharya} \emph
  {et~al.}}]{ALICE:2021gpt}%
  \BibitemOpen
  \bibfield  {author} {\bibinfo {author} {\bibfnamefont {S.}~\bibnamefont
  {Acharya}} \emph {et~al.} (\bibinfo {collaboration} {ALICE}),\ }\bibfield
  {title} {\bibinfo {title} {{Coherent $J/\psi$ and $\psi'$ photoproduction at
  midrapidity in ultra-peripheral Pb-Pb collisions at $\sqrt{s_{\mathrm{NN}}}$
  = 5.02 TeV}},\ }\href {https://doi.org/10.1140/epjc/s10052-021-09437-6}
  {\bibfield  {journal} {\bibinfo  {journal} {Eur. Phys. J. C}\ }\textbf
  {\bibinfo {volume} {81}},\ \bibinfo {pages} {712} (\bibinfo {year}
  {2021}{\natexlab{c}})},\ \Eprint {https://arxiv.org/abs/2101.04577}
  {arXiv:2101.04577 [nucl-ex]} \BibitemShut {NoStop}%
\bibitem [{\citenamefont {Aaij}\ \emph {et~al.}(2022)\citenamefont {Aaij} \emph
  {et~al.}}]{LHCb:2021hoq}%
  \BibitemOpen
  \bibfield  {author} {\bibinfo {author} {\bibfnamefont {R.}~\bibnamefont
  {Aaij}} \emph {et~al.} (\bibinfo {collaboration} {LHCb}),\ }\bibfield
  {title} {\bibinfo {title} {{$J/\psi$ photoproduction in Pb-Pb peripheral
  collisions at $\sqrt{s_{\mathrm{NN}}}$ = 5 TeV}},\ }\href
  {https://doi.org/10.1103/PhysRevC.105.L032201} {\bibfield  {journal}
  {\bibinfo  {journal} {Phys. Rev. C}\ }\textbf {\bibinfo {volume} {105}},\
  \bibinfo {pages} {L032201} (\bibinfo {year} {2022})},\ \Eprint
  {https://arxiv.org/abs/2108.02681} {arXiv:2108.02681 [hep-ex]} \BibitemShut
  {NoStop}%
\bibitem [{\citenamefont {Tumasyan}\ \emph {et~al.}(2023)\citenamefont
  {Tumasyan} \emph {et~al.}}]{CMS:2023snh}%
  \BibitemOpen
  \bibfield  {author} {\bibinfo {author} {\bibfnamefont {A.}~\bibnamefont
  {Tumasyan}} \emph {et~al.} (\bibinfo {collaboration} {CMS}),\ }\bibfield
  {title} {\bibinfo {title} {{Probing small Bjorken-$x$ nuclear gluonic
  structure via coherent J/$\psi$ photoproduction in ultraperipheral PbPb
  collisions at $\sqrt{s_{\mathrm{NN}}}$ = 5.02 TeV}},\ }\href@noop {} {\
  (\bibinfo {year} {2023})},\ \Eprint {https://arxiv.org/abs/2303.16984}
  {arXiv:2303.16984 [nucl-ex]} \BibitemShut {NoStop}%
\bibitem [{\citenamefont {Acharya}\ \emph
  {et~al.}(2023{\natexlab{a}})\citenamefont {Acharya} \emph
  {et~al.}}]{ALICE:2023jgu}%
  \BibitemOpen
  \bibfield  {author} {\bibinfo {author} {\bibfnamefont {S.}~\bibnamefont
  {Acharya}} \emph {et~al.} (\bibinfo {collaboration} {ALICE}),\ }\bibfield
  {title} {\bibinfo {title} {{Energy dependence of coherent photonuclear
  production of J/\ensuremath{\psi} mesons in ultra-peripheral Pb-Pb collisions
  at $ \sqrt{{\textrm{s}}_{\textrm{NN}}} $ = 5.02 TeV}},\ }\href
  {https://doi.org/10.1007/JHEP10(2023)119} {\bibfield  {journal} {\bibinfo
  {journal} {JHEP (2023)}\ }\textbf {\bibinfo {volume} {10}},\ \bibinfo {pages}
  {119} (\bibinfo {year} {2023}{\natexlab{a}})},\ \Eprint
  {https://arxiv.org/abs/2305.19060} {arXiv:2305.19060 [nucl-ex]} \BibitemShut
  {NoStop}%
\bibitem [{\citenamefont {Abelev}\ \emph
  {et~al.}(2013{\natexlab{b}})\citenamefont {Abelev} \emph
  {et~al.}}]{ALICE:2012yye}%
  \BibitemOpen
  \bibfield  {author} {\bibinfo {author} {\bibfnamefont {B.}~\bibnamefont
  {Abelev}} \emph {et~al.} (\bibinfo {collaboration} {ALICE}),\ }\bibfield
  {title} {\bibinfo {title} {{Coherent $J/\psi$ photoproduction in
  ultra-peripheral Pb-Pb collisions at $\sqrt{s_{\mathrm{NN}}} = 2.76$ TeV}},\
  }\href {https://doi.org/10.1016/j.physletb.2012.11.059} {\bibfield  {journal}
  {\bibinfo  {journal} {Phys. Lett. B}\ }\textbf {\bibinfo {volume} {718}},\
  \bibinfo {pages} {1273} (\bibinfo {year} {2013}{\natexlab{b}})},\ \Eprint
  {https://arxiv.org/abs/1209.3715} {arXiv:1209.3715 [nucl-ex]} \BibitemShut
  {NoStop}%
\bibitem [{\citenamefont {Alvioli}\ \emph {et~al.}(2018)\citenamefont
  {Alvioli}, \citenamefont {Frankfurt}, \citenamefont {Guzey}, \citenamefont
  {Strikman},\ and\ \citenamefont {Zhalov}}]{Strikman:2018mbu}%
  \BibitemOpen
  \bibfield  {author} {\bibinfo {author} {\bibfnamefont {M.}~\bibnamefont
  {Alvioli}}, \bibinfo {author} {\bibfnamefont {L.}~\bibnamefont {Frankfurt}},
  \bibinfo {author} {\bibfnamefont {V.}~\bibnamefont {Guzey}}, \bibinfo
  {author} {\bibfnamefont {M.}~\bibnamefont {Strikman}},\ and\ \bibinfo
  {author} {\bibfnamefont {M.}~\bibnamefont {Zhalov}},\ }\bibfield  {title}
  {\bibinfo {title} {{Color fluctuation phenomena in $\gamma A$ collisions at
  the LHC}},\ }\href {https://doi.org/10.23727/CERN-Proceedings-2018-001.151}
  {\bibfield  {journal} {\bibinfo  {journal} {CERN Proc.}\ }\textbf {\bibinfo
  {volume} {1}},\ \bibinfo {pages} {151} (\bibinfo {year} {2018})}\BibitemShut
  {NoStop}%
\bibitem [{\citenamefont {Guzey}\ and\ \citenamefont
  {Zhalov}(2013)}]{Guzey:2013qza}%
  \BibitemOpen
  \bibfield  {author} {\bibinfo {author} {\bibfnamefont {V.}~\bibnamefont
  {Guzey}}\ and\ \bibinfo {author} {\bibfnamefont {M.}~\bibnamefont {Zhalov}},\
  }\bibfield  {title} {\bibinfo {title} {{Exclusive $J/{\psi}$ production in
  ultraperipheral collisions at the LHC: constrains on the gluon distributions
  in the proton and nuclei}},\ }\href {https://doi.org/10.1007/JHEP10(2013)207}
  {\bibfield  {journal} {\bibinfo  {journal} {JHEP}\ }\textbf {\bibinfo
  {volume} {10}},\ \bibinfo {pages} {207 (2013)}},\ \Eprint
  {https://arxiv.org/abs/1307.4526} {arXiv:1307.4526 [hep-ph]} \BibitemShut
  {NoStop}%
\bibitem [{\citenamefont {Guzey}\ \emph {et~al.}(2019)\citenamefont {Guzey},
  \citenamefont {Strikman},\ and\ \citenamefont {Zhalov}}]{Guzey:2018tlk}%
  \BibitemOpen
  \bibfield  {author} {\bibinfo {author} {\bibfnamefont {V.}~\bibnamefont
  {Guzey}}, \bibinfo {author} {\bibfnamefont {M.}~\bibnamefont {Strikman}},\
  and\ \bibinfo {author} {\bibfnamefont {M.}~\bibnamefont {Zhalov}},\
  }\bibfield  {title} {\bibinfo {title} {{Nucleon dissociation and incoherent
  $J/\psi$ photoproduction on nuclei in ion ultraperipheral collisions at the
  Large Hadron Collider}},\ }\href {https://doi.org/10.1103/PhysRevC.99.015201}
  {\bibfield  {journal} {\bibinfo  {journal} {Phys. Rev. C}\ }\textbf {\bibinfo
  {volume} {99}},\ \bibinfo {pages} {015201} (\bibinfo {year} {2019})},\
  \Eprint {https://arxiv.org/abs/1808.00740} {arXiv:1808.00740 [hep-ph]}
  \BibitemShut {NoStop}%
\bibitem [{\citenamefont {M\"antysaari}\ \emph {et~al.}(2022)\citenamefont
  {M\"antysaari}, \citenamefont {Salazar},\ and\ \citenamefont
  {Schenke}}]{Mantysaari:2022sux}%
  \BibitemOpen
  \bibfield  {author} {\bibinfo {author} {\bibfnamefont {H.}~\bibnamefont
  {M\"antysaari}}, \bibinfo {author} {\bibfnamefont {F.}~\bibnamefont
  {Salazar}},\ and\ \bibinfo {author} {\bibfnamefont {B.}~\bibnamefont
  {Schenke}},\ }\bibfield  {title} {\bibinfo {title} {{Nuclear geometry at high
  energy from exclusive vector meson production}},\ }\href@noop {} {\
  (\bibinfo {year} {2022})},\ \Eprint {https://arxiv.org/abs/2207.03712}
  {arXiv:2207.03712 [hep-ph]} \BibitemShut {NoStop}%
\bibitem [{\citenamefont {Sambasivam}\ \emph {et~al.}(2020)\citenamefont
  {Sambasivam}, \citenamefont {Toll},\ and\ \citenamefont
  {Ullrich}}]{Sambasivam:2019gdd}%
  \BibitemOpen
  \bibfield  {author} {\bibinfo {author} {\bibfnamefont {B.}~\bibnamefont
  {Sambasivam}}, \bibinfo {author} {\bibfnamefont {T.}~\bibnamefont {Toll}},\
  and\ \bibinfo {author} {\bibfnamefont {T.}~\bibnamefont {Ullrich}},\
  }\bibfield  {title} {\bibinfo {title} {{Investigating saturation effects in
  ultraperipheral collisions at the LHC with the color dipole model}},\ }\href
  {https://doi.org/10.1016/j.physletb.2020.135277} {\bibfield  {journal}
  {\bibinfo  {journal} {Phys. Lett. B}\ }\textbf {\bibinfo {volume} {803}},\
  \bibinfo {pages} {135277} (\bibinfo {year} {2020})},\ \Eprint
  {https://arxiv.org/abs/1910.02899} {arXiv:1910.02899 [hep-ph]} \BibitemShut
  {NoStop}%
\bibitem [{\citenamefont {Toll}\ and\ \citenamefont
  {Ullrich}(2013)}]{Toll:2012mb}%
  \BibitemOpen
  \bibfield  {author} {\bibinfo {author} {\bibfnamefont {T.}~\bibnamefont
  {Toll}}\ and\ \bibinfo {author} {\bibfnamefont {T.}~\bibnamefont {Ullrich}},\
  }\bibfield  {title} {\bibinfo {title} {{Exclusive diffractive processes in
  electron-ion collisions}},\ }\href
  {https://doi.org/10.1103/PhysRevC.87.024913} {\bibfield  {journal} {\bibinfo
  {journal} {Phys. Rev. C}\ }\textbf {\bibinfo {volume} {87}},\ \bibinfo
  {pages} {024913} (\bibinfo {year} {2013})},\ \Eprint
  {https://arxiv.org/abs/1211.3048} {arXiv:1211.3048 [hep-ph]} \BibitemShut
  {NoStop}%
\bibitem [{\citenamefont {Armesto}(2018)}]{Armesto:2018ljh}%
  \BibitemOpen
  \bibfield  {author} {\bibinfo {author} {\bibfnamefont {N.}~\bibnamefont
  {Armesto}},\ }\bibfield  {title} {\bibinfo {title} {{Small collision systems:
  Theory overview on cold nuclear matter effects}},\ }\href
  {https://doi.org/10.1051/epjconf/201817111001} {\bibfield  {journal}
  {\bibinfo  {journal} {EPJ Web Conf.}\ }\textbf {\bibinfo {volume} {171}},\
  \bibinfo {pages} {11001} (\bibinfo {year} {2018})}\BibitemShut {NoStop}%
\bibitem [{\citenamefont {Sj\"ostrand}(2018)}]{Sjostrand:2017cdm}%
  \BibitemOpen
  \bibfield  {author} {\bibinfo {author} {\bibfnamefont {T.}~\bibnamefont
  {Sj\"ostrand}},\ }\bibfield  {title} {\bibinfo {title} {{The Development of
  MPI Modeling in Pythia}},\ }\href
  {https://doi.org/10.1142/9789813227767_0010} {\bibfield  {journal} {\bibinfo
  {journal} {Adv. Ser. Direct. High Energy Phys.}\ }\textbf {\bibinfo {volume}
  {29}},\ \bibinfo {pages} {191} (\bibinfo {year} {2018})},\ \Eprint
  {https://arxiv.org/abs/1706.02166} {arXiv:1706.02166 [hep-ph]} \BibitemShut
  {NoStop}%
\bibitem [{\citenamefont {Arleo}\ \emph {et~al.}(2022)\citenamefont {Arleo},
  \citenamefont {Jackson},\ and\ \citenamefont {Peign\'e}}]{Arleo:2021bpv}%
  \BibitemOpen
  \bibfield  {author} {\bibinfo {author} {\bibfnamefont {F.}~\bibnamefont
  {Arleo}}, \bibinfo {author} {\bibfnamefont {G.}~\bibnamefont {Jackson}},\
  and\ \bibinfo {author} {\bibfnamefont {S.}~\bibnamefont {Peign\'e}},\
  }\bibfield  {title} {\bibinfo {title} {{Impact of fully coherent energy loss
  on heavy meson production in pA collisions}},\ }\href
  {https://doi.org/10.1007/JHEP01(2022)164} {\bibfield  {journal} {\bibinfo
  {journal} {JHEP}\ }\textbf {\bibinfo {volume} {01}},\ \bibinfo {pages}
  {164}},\ \Eprint {https://arxiv.org/abs/2107.05871} {arXiv:2107.05871
  [hep-ph]} \BibitemShut {NoStop}%
\bibitem [{Note1()}]{Note1}%
  \BibitemOpen
  \bibinfo {note} {Note that there is a $-$0.5 GeV shift in the estimate of
  $\left \langle W_{\protect \mathrm {\gamma *N}}\right \rangle ~$at
  midrapidity $|y|<0.2$, caused by the higher photon flux of the lower energy
  photon contribution; however, the effect of this shift is found to be
  negligible.}\BibitemShut {Stop}%
\bibitem [{\citenamefont {Kryshen}\ \emph {et~al.}(2023)\citenamefont
  {Kryshen}, \citenamefont {Strikman},\ and\ \citenamefont
  {Zhalov}}]{Kryshen:2023bxy}%
  \BibitemOpen
  \bibfield  {author} {\bibinfo {author} {\bibfnamefont {E.}~\bibnamefont
  {Kryshen}}, \bibinfo {author} {\bibfnamefont {M.}~\bibnamefont {Strikman}},\
  and\ \bibinfo {author} {\bibfnamefont {M.}~\bibnamefont {Zhalov}},\
  }\bibfield  {title} {\bibinfo {title} {{Photoproduction of $J/\psi$ with
  neutron tagging in ultra-peripheral collisions of nuclei at RHIC and the
  LHC}},\ }\href@noop {} {\  (\bibinfo {year} {2023})},\ \Eprint
  {https://arxiv.org/abs/2303.12052} {arXiv:2303.12052 [hep-ph]} \BibitemShut
  {NoStop}%
\bibitem [{\citenamefont {Guzey}\ \emph {et~al.}(2013)\citenamefont {Guzey},
  \citenamefont {Kryshen}, \citenamefont {Strikman},\ and\ \citenamefont
  {Zhalov}}]{Guzey:2013xba}%
  \BibitemOpen
  \bibfield  {author} {\bibinfo {author} {\bibfnamefont {V.}~\bibnamefont
  {Guzey}}, \bibinfo {author} {\bibfnamefont {E.}~\bibnamefont {Kryshen}},
  \bibinfo {author} {\bibfnamefont {M.}~\bibnamefont {Strikman}},\ and\
  \bibinfo {author} {\bibfnamefont {M.}~\bibnamefont {Zhalov}},\ }\bibfield
  {title} {\bibinfo {title} {{Evidence for nuclear gluon shadowing from the
  ALICE measurements of PbPb ultraperipheral exclusive $J/{\psi}$
  production}},\ }\href {https://doi.org/10.1016/j.physletb.2013.08.043}
  {\bibfield  {journal} {\bibinfo  {journal} {Phys. Lett. B}\ }\textbf
  {\bibinfo {volume} {726}},\ \bibinfo {pages} {290} (\bibinfo {year}
  {2013})},\ \Eprint {https://arxiv.org/abs/1305.1724} {arXiv:1305.1724
  [hep-ph]} \BibitemShut {NoStop}%
\bibitem [{\citenamefont {M\"antysaari}\ and\ \citenamefont
  {Schenke}(2016)}]{Mantysaari:2016ykx}%
  \BibitemOpen
  \bibfield  {author} {\bibinfo {author} {\bibfnamefont {H.}~\bibnamefont
  {M\"antysaari}}\ and\ \bibinfo {author} {\bibfnamefont {B.}~\bibnamefont
  {Schenke}},\ }\bibfield  {title} {\bibinfo {title} {{Evidence of strong
  proton shape fluctuations from incoherent diffraction}},\ }\href
  {https://doi.org/10.1103/PhysRevLett.117.052301} {\bibfield  {journal}
  {\bibinfo  {journal} {Phys. Rev. Lett.}\ }\textbf {\bibinfo {volume} {117}},\
  \bibinfo {pages} {052301} (\bibinfo {year} {2016})},\ \Eprint
  {https://arxiv.org/abs/1603.04349} {arXiv:1603.04349 [hep-ph]} \BibitemShut
  {NoStop}%
\bibitem [{\citenamefont {Ackermann}\ \emph {et~al.}(2003)\citenamefont
  {Ackermann} \emph {et~al.}}]{Ackermann:2002ad}%
  \BibitemOpen
  \bibfield  {author} {\bibinfo {author} {\bibfnamefont {K.~H.}\ \bibnamefont
  {Ackermann}} \emph {et~al.} (\bibinfo {collaboration} {STAR}),\ }\bibfield
  {title} {\bibinfo {title} {{STAR detector overview}},\ }\href
  {https://doi.org/10.1016/S0168-9002(02)01960-5} {\bibfield  {journal}
  {\bibinfo  {journal} {Nucl. Instrum. Meth. A}\ }\textbf {\bibinfo {volume}
  {499}},\ \bibinfo {pages} {624} (\bibinfo {year} {2003})}\BibitemShut
  {NoStop}%
\bibitem [{\citenamefont {Adam}\ \emph {et~al.}(2018)\citenamefont {Adam} \emph
  {et~al.}}]{Adam:2018tdm}%
  \BibitemOpen
  \bibfield  {author} {\bibinfo {author} {\bibfnamefont {J.}~\bibnamefont
  {Adam}} \emph {et~al.} (\bibinfo {collaboration} {STAR}),\ }\bibfield
  {title} {\bibinfo {title} {{Low-$p_T$ $e^{+}e^{-}$ pair production in Au$+$Au
  collisions at $\sqrt{s_{\mathrm{NN}}}$ = 200 GeV and U$+$U collisions at
  $\sqrt{s_{\mathrm{NN}}}$ = 193 GeV at STAR}},\ }\href
  {https://doi.org/10.1103/PhysRevLett.121.132301} {\bibfield  {journal}
  {\bibinfo  {journal} {Phys. Rev. Lett.}\ }\textbf {\bibinfo {volume} {121}},\
  \bibinfo {pages} {132301} (\bibinfo {year} {2018})},\ \Eprint
  {https://arxiv.org/abs/1806.02295} {arXiv:1806.02295 [hep-ex]} \BibitemShut
  {NoStop}%
\bibitem [{\citenamefont {Adam}\ \emph {et~al.}(2021)\citenamefont {Adam} \emph
  {et~al.}}]{Adam:2020cwy}%
  \BibitemOpen
  \bibfield  {author} {\bibinfo {author} {\bibfnamefont {J.}~\bibnamefont
  {Adam}} \emph {et~al.} (\bibinfo {collaboration} {STAR}),\ }\bibfield
  {title} {\bibinfo {title} {{Measurements of $W$ and $Z/\gamma^*$ cross
  sections and their ratios in p+p collisions at RHIC}},\ }\href
  {https://doi.org/10.1103/PhysRevD.103.012001} {\bibfield  {journal} {\bibinfo
   {journal} {Phys. Rev. D}\ }\textbf {\bibinfo {volume} {103}},\ \bibinfo
  {pages} {012001} (\bibinfo {year} {2021})},\ \Eprint
  {https://arxiv.org/abs/2011.04708} {arXiv:2011.04708 [nucl-ex]} \BibitemShut
  {NoStop}%
\bibitem [{\citenamefont {Anderson}\ \emph {et~al.}(2003)\citenamefont
  {Anderson} \emph {et~al.}}]{Anderson:2003ur}%
  \BibitemOpen
  \bibfield  {author} {\bibinfo {author} {\bibfnamefont {M.}~\bibnamefont
  {Anderson}} \emph {et~al.},\ }\bibfield  {title} {\bibinfo {title} {{The Star
  time projection chamber: A Unique tool for studying high multiplicity events
  at RHIC}},\ }\href {https://doi.org/10.1016/S0168-9002(02)01964-2} {\bibfield
   {journal} {\bibinfo  {journal} {Nucl. Instrum. Meth. A}\ }\textbf {\bibinfo
  {volume} {499}},\ \bibinfo {pages} {659} (\bibinfo {year} {2003})},\ \Eprint
  {https://arxiv.org/abs/nucl-ex/0301015} {arXiv:nucl-ex/0301015} \BibitemShut
  {NoStop}%
\bibitem [{\citenamefont {Beddo}\ \emph {et~al.}(2003)\citenamefont {Beddo}
  \emph {et~al.}}]{Beddo:2002zx}%
  \BibitemOpen
  \bibfield  {author} {\bibinfo {author} {\bibfnamefont {M.}~\bibnamefont
  {Beddo}} \emph {et~al.} (\bibinfo {collaboration} {STAR}),\ }\bibfield
  {title} {\bibinfo {title} {{The STAR barrel electromagnetic calorimeter}},\
  }\href {https://doi.org/10.1016/S0168-9002(02)01970-8} {\bibfield  {journal}
  {\bibinfo  {journal} {Nucl. Instrum. Meth. A}\ }\textbf {\bibinfo {volume}
  {499}},\ \bibinfo {pages} {725} (\bibinfo {year} {2003})}\BibitemShut
  {NoStop}%
\bibitem [{\citenamefont {Whitten}(2008)}]{Whitten:2008zz}%
  \BibitemOpen
  \bibfield  {author} {\bibinfo {author} {\bibfnamefont {C.~A.}\ \bibnamefont
  {Whitten}} (\bibinfo {collaboration} {STAR}),\ }\bibfield  {title} {\bibinfo
  {title} {{The beam-beam counter: A local polarimeter at STAR}},\ }\href
  {https://doi.org/10.1063/1.2888113} {\bibfield  {journal} {\bibinfo
  {journal} {AIP Conf. Proc.}\ }\textbf {\bibinfo {volume} {980}},\ \bibinfo
  {pages} {390} (\bibinfo {year} {2008})}\BibitemShut {NoStop}%
\bibitem [{\citenamefont {Klein}\ \emph {et~al.}(2017)\citenamefont {Klein},
  \citenamefont {Nystrand}, \citenamefont {Seger}, \citenamefont {Gorbunov},\
  and\ \citenamefont {Butterworth}}]{Klein:2016yzr}%
  \BibitemOpen
  \bibfield  {author} {\bibinfo {author} {\bibfnamefont {S.~R.}\ \bibnamefont
  {Klein}}, \bibinfo {author} {\bibfnamefont {J.}~\bibnamefont {Nystrand}},
  \bibinfo {author} {\bibfnamefont {J.}~\bibnamefont {Seger}}, \bibinfo
  {author} {\bibfnamefont {Y.}~\bibnamefont {Gorbunov}},\ and\ \bibinfo
  {author} {\bibfnamefont {J.}~\bibnamefont {Butterworth}},\ }\bibfield
  {title} {\bibinfo {title} {{STARlight: A Monte Carlo simulation program for
  ultra-peripheral collisions of relativistic ions}},\ }\href
  {https://doi.org/10.1016/j.cpc.2016.10.016} {\bibfield  {journal} {\bibinfo
  {journal} {Comput. Phys. Commun.}\ }\textbf {\bibinfo {volume} {212}},\
  \bibinfo {pages} {258} (\bibinfo {year} {2017})},\ \Eprint
  {https://arxiv.org/abs/1607.03838} {arXiv:1607.03838 [hep-ph]} \BibitemShut
  {NoStop}%
\bibitem [{\citenamefont {Alexa}\ \emph {et~al.}(2013)\citenamefont {Alexa}
  \emph {et~al.}}]{Alexa:2013xxa}%
  \BibitemOpen
  \bibfield  {author} {\bibinfo {author} {\bibfnamefont {C.}~\bibnamefont
  {Alexa}} \emph {et~al.} (\bibinfo {collaboration} {H1}),\ }\bibfield  {title}
  {\bibinfo {title} {{Elastic and Proton-Dissociative Photoproduction of
  J/$\psi$ Mesons at HERA}},\ }\href
  {https://doi.org/10.1140/epjc/s10052-013-2466-y} {\bibfield  {journal}
  {\bibinfo  {journal} {Eur. Phys. J. C}\ }\textbf {\bibinfo {volume} {73}},\
  \bibinfo {pages} {2466} (\bibinfo {year} {2013})},\ \Eprint
  {https://arxiv.org/abs/1304.5162} {arXiv:1304.5162 [hep-ex]} \BibitemShut
  {NoStop}%
\bibitem [{\citenamefont {Brun}\ \emph {et~al.}(1987)\citenamefont {Brun},
  \citenamefont {Bruyant}, \citenamefont {Maire}, \citenamefont {McPherson},\
  and\ \citenamefont {Zanarini}}]{Brun:1987ma}%
  \BibitemOpen
  \bibfield  {author} {\bibinfo {author} {\bibfnamefont {R.}~\bibnamefont
  {Brun}}, \bibinfo {author} {\bibfnamefont {F.}~\bibnamefont {Bruyant}},
  \bibinfo {author} {\bibfnamefont {M.}~\bibnamefont {Maire}}, \bibinfo
  {author} {\bibfnamefont {A.~C.}\ \bibnamefont {McPherson}},\ and\ \bibinfo
  {author} {\bibfnamefont {P.}~\bibnamefont {Zanarini}},\ }\href@noop {}
  {\bibinfo {title} {{GEANT3, CERN-DD-EE-84-01}}},\ \bibinfo {howpublished}
  {\url{https://cds.cern.ch/record/1119728}} (\bibinfo {year}
  {1987})\BibitemShut {NoStop}%
\bibitem [{STA(2023)}]{STAR:2023gpk}%
  \BibitemOpen
  \bibfield  {title} {\bibinfo {title} {{Exclusive $J/\psi$, $\psi(2s)$, and
  $e^{+}e^{-}$ pair production in Au$+$Au ultra-peripheral collisions at
  RHIC}},\ }\href@noop {} {\  (\bibinfo {year} {2023})},\ \Eprint
  {https://arxiv.org/abs/2311.13632} {arXiv:2311.13632 [nucl-ex]} \BibitemShut
  {NoStop}%
\bibitem [{\citenamefont {Zyla}\ \emph {et~al.}(2020)\citenamefont {Zyla} \emph
  {et~al.}}]{ParticleDataGroup:2020ssz}%
  \BibitemOpen
  \bibfield  {author} {\bibinfo {author} {\bibfnamefont {P.~A.}\ \bibnamefont
  {Zyla}} \emph {et~al.} (\bibinfo {collaboration} {Particle Data Group}),\
  }\bibfield  {title} {\bibinfo {title} {{Review of Particle Physics}},\ }\href
  {https://doi.org/10.1093/ptep/ptaa104} {\bibfield  {journal} {\bibinfo
  {journal} {PTEP}\ }\textbf {\bibinfo {volume} {2020}},\ \bibinfo {pages}
  {083C01} (\bibinfo {year} {2020})}\BibitemShut {NoStop}%
\bibitem [{\citenamefont {Guzey}\ \emph {et~al.}(2014)\citenamefont {Guzey},
  \citenamefont {Strikman},\ and\ \citenamefont {Zhalov}}]{Guzey:2013jaa}%
  \BibitemOpen
  \bibfield  {author} {\bibinfo {author} {\bibfnamefont {V.}~\bibnamefont
  {Guzey}}, \bibinfo {author} {\bibfnamefont {M.}~\bibnamefont {Strikman}},\
  and\ \bibinfo {author} {\bibfnamefont {M.}~\bibnamefont {Zhalov}},\
  }\bibfield  {title} {\bibinfo {title} {{Disentangling coherent and incoherent
  quasielastic $J/\psi$ photoproduction on nuclei by neutron tagging in
  ultraperipheral ion collisions at the LHC}},\ }\href
  {https://doi.org/10.1140/epjc/s10052-014-2942-z} {\bibfield  {journal}
  {\bibinfo  {journal} {Eur. Phys. J. C}\ }\textbf {\bibinfo {volume} {74}},\
  \bibinfo {pages} {2942} (\bibinfo {year} {2014})},\ \Eprint
  {https://arxiv.org/abs/1312.6486} {arXiv:1312.6486 [hep-ph]} \BibitemShut
  {NoStop}%
\bibitem [{\citenamefont {Abdallah}\ \emph {et~al.}(2022)\citenamefont
  {Abdallah} \emph {et~al.}}]{STAR:2021wwq}%
  \BibitemOpen
  \bibfield  {author} {\bibinfo {author} {\bibfnamefont {M.}~\bibnamefont
  {Abdallah}} \emph {et~al.} (\bibinfo {collaboration} {STAR}),\ }\bibfield
  {title} {\bibinfo {title} {{Probing the Gluonic Structure of the Deuteron
  with $J/\psi$ Photoproduction in d+Au Ultraperipheral Collisions}},\ }\href
  {https://doi.org/10.1103/PhysRevLett.128.122303} {\bibfield  {journal}
  {\bibinfo  {journal} {Phys. Rev. Lett.}\ }\textbf {\bibinfo {volume} {128}},\
  \bibinfo {pages} {122303} (\bibinfo {year} {2022})},\ \Eprint
  {https://arxiv.org/abs/2109.07625} {arXiv:2109.07625 [nucl-ex]} \BibitemShut
  {NoStop}%
\bibitem [{\citenamefont {Chang}\ \emph {et~al.}(2021)\citenamefont {Chang},
  \citenamefont {Aschenauer}, \citenamefont {Baker}, \citenamefont {Jentsch},
  \citenamefont {Lee}, \citenamefont {Tu}, \citenamefont {Yin},\ and\
  \citenamefont {Zheng}}]{Chang:2021jnu}%
  \BibitemOpen
  \bibfield  {author} {\bibinfo {author} {\bibfnamefont {W.}~\bibnamefont
  {Chang}}, \bibinfo {author} {\bibfnamefont {E.-C.}\ \bibnamefont
  {Aschenauer}}, \bibinfo {author} {\bibfnamefont {M.~D.}\ \bibnamefont
  {Baker}}, \bibinfo {author} {\bibfnamefont {A.}~\bibnamefont {Jentsch}},
  \bibinfo {author} {\bibfnamefont {J.-H.}\ \bibnamefont {Lee}}, \bibinfo
  {author} {\bibfnamefont {Z.}~\bibnamefont {Tu}}, \bibinfo {author}
  {\bibfnamefont {Z.}~\bibnamefont {Yin}},\ and\ \bibinfo {author}
  {\bibfnamefont {L.}~\bibnamefont {Zheng}},\ }\bibfield  {title} {\bibinfo
  {title} {{Investigation of the background in coherent J/\ensuremath{\psi}
  production at the EIC}},\ }\href
  {https://doi.org/10.1103/PhysRevD.104.114030} {\bibfield  {journal} {\bibinfo
   {journal} {Phys. Rev. D}\ }\textbf {\bibinfo {volume} {104}},\ \bibinfo
  {pages} {114030} (\bibinfo {year} {2021})},\ \Eprint
  {https://arxiv.org/abs/2108.01694} {arXiv:2108.01694 [nucl-ex]} \BibitemShut
  {NoStop}%
\bibitem [{\citenamefont {Eskola}\ \emph {et~al.}(2023)\citenamefont {Eskola},
  \citenamefont {Flett}, \citenamefont {Guzey}, \citenamefont {L\"oyt\"ainen},\
  and\ \citenamefont {Paukkunen}}]{Eskola:2022vaf}%
  \BibitemOpen
  \bibfield  {author} {\bibinfo {author} {\bibfnamefont {K.~J.}\ \bibnamefont
  {Eskola}}, \bibinfo {author} {\bibfnamefont {C.~A.}\ \bibnamefont {Flett}},
  \bibinfo {author} {\bibfnamefont {V.}~\bibnamefont {Guzey}}, \bibinfo
  {author} {\bibfnamefont {T.}~\bibnamefont {L\"oyt\"ainen}},\ and\ \bibinfo
  {author} {\bibfnamefont {H.}~\bibnamefont {Paukkunen}},\ }\bibfield  {title}
  {\bibinfo {title} {{Next-to-leading order perturbative QCD predictions for
  exclusive $J/\psi$ photoproduction in oxygen-oxygen and lead-lead collisions
  at energies available at the CERN Large Hadron Collider}},\ }\href
  {https://doi.org/10.1103/PhysRevC.107.044912} {\bibfield  {journal} {\bibinfo
   {journal} {Phys. Rev. C}\ }\textbf {\bibinfo {volume} {107}},\ \bibinfo
  {pages} {044912} (\bibinfo {year} {2023})},\ \Eprint
  {https://arxiv.org/abs/2210.16048} {arXiv:2210.16048 [hep-ph]} \BibitemShut
  {NoStop}%
\bibitem [{\citenamefont {Eskola}\ \emph
  {et~al.}(2022{\natexlab{a}})\citenamefont {Eskola}, \citenamefont {Flett},
  \citenamefont {Guzey}, \citenamefont {L\"oyt\"ainen},\ and\ \citenamefont
  {Paukkunen}}]{Eskola:2022vpi}%
  \BibitemOpen
  \bibfield  {author} {\bibinfo {author} {\bibfnamefont {K.~J.}\ \bibnamefont
  {Eskola}}, \bibinfo {author} {\bibfnamefont {C.~A.}\ \bibnamefont {Flett}},
  \bibinfo {author} {\bibfnamefont {V.}~\bibnamefont {Guzey}}, \bibinfo
  {author} {\bibfnamefont {T.}~\bibnamefont {L\"oyt\"ainen}},\ and\ \bibinfo
  {author} {\bibfnamefont {H.}~\bibnamefont {Paukkunen}},\ }\bibfield  {title}
  {\bibinfo {title} {{Exclusive J/\ensuremath{\psi} photoproduction in
  ultraperipheral Pb+Pb collisions at the CERN Large Hadron Collider calculated
  at next-to-leading order perturbative QCD}},\ }\href
  {https://doi.org/10.1103/PhysRevC.106.035202} {\bibfield  {journal} {\bibinfo
   {journal} {Phys. Rev. C}\ }\textbf {\bibinfo {volume} {106}},\ \bibinfo
  {pages} {035202} (\bibinfo {year} {2022}{\natexlab{a}})},\ \Eprint
  {https://arxiv.org/abs/2203.11613} {arXiv:2203.11613 [hep-ph]} \BibitemShut
  {NoStop}%
\bibitem [{\citenamefont {Ducati}\ \emph {et~al.}(2013)\citenamefont {Ducati},
  \citenamefont {Griep},\ and\ \citenamefont {Machado}}]{Ducati:2013bya}%
  \BibitemOpen
  \bibfield  {author} {\bibinfo {author} {\bibfnamefont {M.~B.~G.}\
  \bibnamefont {Ducati}}, \bibinfo {author} {\bibfnamefont {M.~T.}\
  \bibnamefont {Griep}},\ and\ \bibinfo {author} {\bibfnamefont {M.~V.~T.}\
  \bibnamefont {Machado}},\ }\bibfield  {title} {\bibinfo {title} {{Diffractive
  photoproduction of radially excited psi(2S) mesons in photon-Pomeron
  reactions in PbPb collisions at the CERN LHC}},\ }\href
  {https://doi.org/10.1103/PhysRevC.88.014910} {\bibfield  {journal} {\bibinfo
  {journal} {Phys. Rev. C}\ }\textbf {\bibinfo {volume} {88}},\ \bibinfo
  {pages} {014910} (\bibinfo {year} {2013})},\ \Eprint
  {https://arxiv.org/abs/1305.2407} {arXiv:1305.2407 [hep-ph]} \BibitemShut
  {NoStop}%
\bibitem [{\citenamefont {Adloff}\ \emph {et~al.}(2002)\citenamefont {Adloff}
  \emph {et~al.}}]{H1:2002yab}%
  \BibitemOpen
  \bibfield  {author} {\bibinfo {author} {\bibfnamefont {C.}~\bibnamefont
  {Adloff}} \emph {et~al.} (\bibinfo {collaboration} {H1}),\ }\bibfield
  {title} {\bibinfo {title} {{Diffractive photoproduction of psi(2S) mesons at
  HERA}},\ }\href {https://doi.org/10.1016/S0370-2693(02)02275-X} {\bibfield
  {journal} {\bibinfo  {journal} {Phys. Lett. B}\ }\textbf {\bibinfo {volume}
  {541}},\ \bibinfo {pages} {251} (\bibinfo {year} {2002})},\ \Eprint
  {https://arxiv.org/abs/hep-ex/0205107} {arXiv:hep-ex/0205107} \BibitemShut
  {NoStop}%
\bibitem [{\citenamefont {Aaltonen}\ \emph {et~al.}(2009)\citenamefont
  {Aaltonen} \emph {et~al.}}]{CDF:2009xey}%
  \BibitemOpen
  \bibfield  {author} {\bibinfo {author} {\bibfnamefont {T.}~\bibnamefont
  {Aaltonen}} \emph {et~al.} (\bibinfo {collaboration} {CDF}),\ }\bibfield
  {title} {\bibinfo {title} {{Observation of exclusive charmonium production
  and $\gamma+\gamma$ to $\mu^+\mu^-$ in $p\bar{p}$ collisions at $\sqrt{s} =
  1.96$ TeV}},\ }\href {https://doi.org/10.1103/PhysRevLett.102.242001}
  {\bibfield  {journal} {\bibinfo  {journal} {Phys. Rev. Lett.}\ }\textbf
  {\bibinfo {volume} {102}},\ \bibinfo {pages} {242001} (\bibinfo {year}
  {2009})},\ \Eprint {https://arxiv.org/abs/0902.1271} {arXiv:0902.1271
  [hep-ex]} \BibitemShut {NoStop}%
\bibitem [{\citenamefont {Aaij}\ \emph {et~al.}(2014)\citenamefont {Aaij} \emph
  {et~al.}}]{LHCb:2014acg}%
  \BibitemOpen
  \bibfield  {author} {\bibinfo {author} {\bibfnamefont {R.}~\bibnamefont
  {Aaij}} \emph {et~al.} (\bibinfo {collaboration} {LHCb}),\ }\bibfield
  {title} {\bibinfo {title} {{Updated measurements of exclusive $J/\psi$ and
  $\psi$(2S) production cross-sections in pp collisions at $\sqrt{s}=7$ TeV}},\
  }\href {https://doi.org/10.1088/0954-3899/41/5/055002} {\bibfield  {journal}
  {\bibinfo  {journal} {J. Phys. G}\ }\textbf {\bibinfo {volume} {41}},\
  \bibinfo {pages} {055002} (\bibinfo {year} {2014})},\ \Eprint
  {https://arxiv.org/abs/1401.3288} {arXiv:1401.3288 [hep-ex]} \BibitemShut
  {NoStop}%
\bibitem [{\citenamefont {Eskola}\ \emph
  {et~al.}(2022{\natexlab{b}})\citenamefont {Eskola}, \citenamefont
  {Paakkinen}, \citenamefont {Paukkunen},\ and\ \citenamefont
  {Salgado}}]{Eskola:2021mjl}%
  \BibitemOpen
  \bibfield  {author} {\bibinfo {author} {\bibfnamefont {K.~J.}\ \bibnamefont
  {Eskola}}, \bibinfo {author} {\bibfnamefont {P.}~\bibnamefont {Paakkinen}},
  \bibinfo {author} {\bibfnamefont {H.}~\bibnamefont {Paukkunen}},\ and\
  \bibinfo {author} {\bibfnamefont {C.~A.}\ \bibnamefont {Salgado}},\
  }\bibfield  {title} {\bibinfo {title} {{Towards EPPS21 nuclear PDFs}},\
  }\href {https://doi.org/10.21468/SciPostPhysProc.8.033} {\bibfield  {journal}
  {\bibinfo  {journal} {SciPost Phys. Proc.}\ }\textbf {\bibinfo {volume}
  {8}},\ \bibinfo {pages} {033} (\bibinfo {year} {2022}{\natexlab{b}})},\
  \Eprint {https://arxiv.org/abs/2106.13661} {arXiv:2106.13661 [hep-ph]}
  \BibitemShut {NoStop}%
\bibitem [{\citenamefont {Acharya}\ \emph
  {et~al.}(2023{\natexlab{b}})\citenamefont {Acharya} \emph
  {et~al.}}]{ALICE:2023gcs}%
  \BibitemOpen
  \bibfield  {author} {\bibinfo {author} {\bibfnamefont {S.}~\bibnamefont
  {Acharya}} \emph {et~al.} (\bibinfo {collaboration} {ALICE}),\ }\bibfield
  {title} {\bibinfo {title} {{First measurement of the $|t|$-dependence of
  incoherent J/$\psi$ photonuclear production}},\ }\href@noop {} {\  (\bibinfo
  {year} {2023}{\natexlab{b}})},\ \Eprint {https://arxiv.org/abs/2305.06169}
  {arXiv:2305.06169 [nucl-ex]} \BibitemShut {NoStop}%
\end{thebibliography}%
\end{document}